\documentclass[aps,prl,floats,twocolumn,showpacs,superscriptaddress]{revtex4}

\usepackage{epsfig}
\usepackage{latexsym,amsmath}

\begin{document}

\title{Emergence of clustering, correlations, and communities
  in a social network model}

\author{Mari{\'a}n Bogu{\~n}{\'a}}

\affiliation{Departament de F{\'\i}sica Fonamental, Universitat de
  Barcelona, Av. Diagonal 647, 08028 Barcelona, Spain}

\author{Romualdo Pastor-Satorras}

\affiliation{Departament de F{\'\i}sica i Enginyeria Nuclear, Universitat
  Polit{\`e}cnica de Catalunya, Campus Nord B4, 08034 Barcelona, Spain}

\author{Albert D{\'\i}az-Guilera}

\affiliation{Departament de F{\'\i}sica Fonamental, Universitat de
  Barcelona, Av. Diagonal 647, 08028 Barcelona, Spain}

\author{Alex Arenas}

\affiliation{Departament d'Enginyeria Inform{\`a}tica i Matem{\`a}tiques,
  Universitat Rovira i Virgili, 43007 Tarragona, Spain}

\date{\today}

\begin{abstract}
  We propose a simple model of social network formation that
  parameterizes the tendency to establish acquaintances by the
  relative distance in a representative social space. By means of
  analytical calculations and numerical simulations, we show that the
  model reproduces the main characteristics of real social networks:
  large clustering coefficient, assortative degree correlations, and
  the emergence of a hierarchy of communities. Our results highlight
  the importance of communities in the understanding of the structure
  of social networks.
\end{abstract}


\maketitle

A considerable effort has been devoted in recent years to the
understanding of complex systems that can be described in term of
networks, in which vertices represent interacting units and edges
stand for the presence of interactions between them
\cite{barabasi02,mendesbook}. Examples of this new brand of complex
networks have been found in systems as diverse as the Internet, the
World-Wide-Web, food-webs, and biological and social organizations
(see \cite{barabasi02,mendesbook} and references therein).

While most of these so-called \textit{complex networks} share many
common traits that hint towards the possibility of common underlying
structural principles \cite{barabasi02,mendesbook}, social networks
\cite{wass94} seem to show some essential differences that place them
apart from other technological or biological networks
\cite{newmansocialdiff}. The main differences between social and
non-social networks can be summarized in the following three
properties: (i) \textit{Clustering}: The property of clustering can be
measured by means of the clustering coefficient \cite{watts98},
defined as the probability that a pair of vertices with a common
neighbor are also connected to each other.  While most complex
networks show a quite large level of clustering \cite{barabasi02}, it
has been recently shown that in some cases the value of the clustering
coefficient can be mostly accounted for by a simple random network
model in which edges are placed at random, under the constraint of a
fixed degree distribution $P(k)$ (defined as the probability that a
vertex is connected to $k$ neighbors, i.e. has degree $k$)
\cite{newman01c,newmanrev}.  For networks with a scale-free degree
distribution of the form $P(k) \sim k^{-\gamma}$, this random construction
can yield noticeable values of the clustering coefficient for finite
networks, indicating that, in this case, the clustering could be a
merely topological property.  This construction, however, cannot
explain the large clustering coefficient observed in social networks
with a bounded, non scale-free degree distribution \cite{amaral}.
(ii) \textit{Degree correlations}: It has been recently recognized
\cite{alexei,assortative} that real networks show degree correlations,
in the sense that the degrees at the end points of any given edge are
not independent.  In particular, this feature can be quantitatively
measured by computing the average degree of the nearest neighbors of a
vertex of degree $k$, $\bar{k}_{nn}(k)$ \cite{alexei}. In this sense,
non-social networks exhibit \textit{disassortative mixing}, implying
that highly connected vertices tend to connect to vertices with small
degree, and vice-versa. This property translates in a decreasing
$\bar{k}_{nn}(k)$ function. Social networks, on the other hand,
display a strong \textit{assortative mixing}, with high degree
vertices connecting preferably to highly connected vertices, a fact
that is reflected in an increasing $\bar{k}_{nn}(k)$ function.  It has
been pointed out \cite{maslovcorr,newmansocialdiff} that, for finite
networks, disassortative mixing can be obtained from a purely random
model, by just imposing the condition of having no more than one edge
between vertices. This observation implies that negative correlation
can find a simple structural explanation; explanation that, on the
other hand, does not apply to social networks, which must be driven by
different organizational principles.  (iii) \textit{Community
  structure}: Social networks possess a complex community structure
\cite{girvan02,guimera03,spectrohuberman}, in which individuals
typically belong to groups or communities, with a high density of
internal connections and loosely connected among them, that on their
turn belong to groups of groups and so on, giving raise to a hierarchy
of nested social communities of practice showing in some cases a
self-similar structure \cite{guimera03}.

Several authors \cite{girvan02,newmansocialdiff,guimera03} have
advocated this last property, the presence of a community structure,
as the very distinguishing feature of social networks, responsible for
the rest of the properties that differentiate those from non-social
networks. In this spirit, in the present paper we propose a model of
social networks in which each vertex (individual) has associated a
position in a certain social space \cite{dodds}, whose coordinates
account for the different characteristics that define their relative
social position with respect to the rest of the individuals.
Individuals establish social connections (acquaintances) with a
probability decreasing with their relative social distance (properly
defined in the social space). This property yields as a result the
presence of communities, defined as local clusters of individuals in a
given social space neighborhood. For general forms of the connecting
probability, the model yields networks of acquaintances with a
non-vanishing clustering coefficient in the thermodynamic limit, plus
general assortative correlations. For a certain range of connectivity
probabilities, moreover, the model reproduces a community structure
with self-similar properties. The model we propose resembles the
hierarchical network model proposed in Ref.~\cite{dodds} (see also
\cite{Kleinberg01nips}). Our approach differs, however, in the fact
that hierarchies are not defined \textit{a priori}, but they emerge as
a result of the construction process.

Our model can be described as follows: Let us consider a set of
$N$ disconnected individuals which are randomly placed within a
social space, $\mathcal{H}$, according to the density
$\rho(\vec{h})$, where vector $\vec{h_i}\equiv (h_i^1,
\cdots,h_i^{d_\mathcal{H}})$ defines the position of the $i$-th
individual and $d_\mathcal{H}$ is the dimension of $\mathcal{H}$.
Each subspace of $\mathcal{H}$ (defined by the different
coordinates of the vector $\vec{h}$) represents a distinctive
social feature, such as profession, religion, geographic location,
etc. and, in general, it will be parametrized by means of a
continuous variable with a domain growing with the size of the
population. This choice is justified by the fact there are not two
identical individuals and, thus, increasing the number of
individuals also increases the diversity of the society.  Even
though it is not strictly necessary for our further development,
we also assume that different subspaces are uncorrelated and,
therefore, we can factorize the total density as
$\rho(\vec{h})=\prod_{n=1}^{d_\mathcal{H}} \rho_n(h^n)$. Assuming
again the independence of social subspaces, we assign a connection
probability between any two pairs of individuals, $\vec{h_i}$ and
$\vec{h_j}$, given by
\begin{equation}
  r(\vec{h_i}, \vec{h_j})=\sum_{n=1}^{d_\mathcal{H}} \omega_n r_n(h_i^n,h_j^n)
  \label{rtotal}
\end{equation}
where $\omega_n$ is a normalized weight factor 
measuring the importance
that each social attribute has in the process of formation of
connections. The key point of our model is the concept of social
distance across each subspace \cite{dodds}. We assume that given
two nodes $i$ and $j$ with respective social coordinates
$\vec{h_i}$ and $\vec{h_j}$, it is possible to define a set of
distances corresponding to each subspace, $d_n (h_i^n,h_j^n) \in
[0,\infty)$, $n=1,\cdots d_\mathcal{H}$. Moreover, we expect that
the probability of acquaintance decreases with social distance.
Therefore, we propose a connection probability
\begin{equation}
r_n(h_i^n,h_j^n)=\frac{1}{1+\left[b_n^{-1} d_n (h_i^n,h_j^n)
\right]^{\alpha_n}} \label{r_n}
\end{equation}
where $b_n$ is a characteristic length scale (that, eventually,
will control the average degree) and $\alpha_n > 1$ is a measure
of homophyly \cite{dodds}, that is, the tendency of people to
connect to similar people.

The degree distribution $P(k)$ of the network can be computed using
the conditional probability $g(k | \vec{h})$ (propagator) that an
individual with social coordinates $\vec{h}$ has $k$ connections
\cite{hiddenvars}.  We can thus write $P(k)=\int \rho(\vec{h}) g(k |
\vec{h}) dh$, where $dh$ stands for the measure element of space
$\mathcal{H}$.  The propagator $g(k | \vec{h})$ can be easily computed
using standard techniques of probability theory \cite{hiddenvars},
leading to a binomial distribution
\begin{equation}
  g(k | \vec{h})= \binom{N-1}{k}
  \left(\frac{\bar{k}(\vec{h})}{N-1}\right)^k
  \left(1-\frac{\bar{k}(\vec{h})}{N-1}\right)^{N-1-k}
\end{equation}
where $\bar{k}(\vec{h})$ is the average degree of individuals with
social coordinate $\vec{h}$. For uncorrelated social subspaces, this
average degree takes the form

\begin{equation}
\bar{k}(\vec{h})=(N-1)\sum_{n=1}^{d_\mathcal{H}} \omega_n \int
\rho_n(h'^n) r_n(h^n,h'^n) dh'^n.
\end{equation}

In the case of a sparse network---constant average degree---the
propagator takes a Poisson form \cite{hiddenvars} and the degree
distribution can simply be written as
\begin{equation}
P(k)=\frac{1}{k!} \int \rho(\vec{h}) [\bar{k}(\vec{h})]^k
e^{-\bar{k}(\vec{h})} dh
\end{equation}
Therefore, if the population is homogeneously distributed in the
social space, the degree distribution will be bounded, in agreement
with the observations made in several real social systems
\cite{guimera03,ebel02,amaral} \footnote{This form of the degree
  distribution is due to the particular connection probability
  Eq.~(\ref{r_n}) considered. More complex forms can yield different
  degree distributions, even with scale-free behavior.}.

The clustering coefficient is defined as the probability that two
neighbors of a given individual are also neighbors themselves.
Following \cite{hiddenvars}, we first compute the probability that an
individual with social vector $\vec{h}$ is connected to an individual
with vector $\vec{h'}$, $p(\vec{h'} | \vec{h})$. This probability
reads $p(\vec{h'} | \vec{h}) = (N-1) \rho(\vec{h'})r(\vec{h},\vec{h'}) /
\bar{k}(\vec{h})$.  Given the independent assignment of edges among
individuals, the clustering coefficient of an individual with vector
$\vec{h}$ is
\begin{equation}
c(\vec{h})=\int \int p(\vec{h'} | \vec{h}) r(\vec{h'},\vec{h''})
p(\vec{h''} | \vec{h}) dh' dh'' \label{cluster_h}
\end{equation}
and the average clustering coefficient is simply given by
\begin{equation}
  \langle c \rangle =\int \rho(\vec{h}) c(\vec{h}) dh
\end{equation}

In order to test the behavior of our model, we consider the simplest
case of a single social feature, i.e.  $d_\mathcal{H}=1$.  As we will
see, even in this case our model presents several non-trivial
properties, that are the signature of real social networks.
Considering the space $\mathcal{H}$ to be the one-dimensional segment
$[0,h_{max}]$, we assign individuals a random, uniformly distributed,
position, i.e. $\rho(h) = 1/ h_{max}$. In this way, the density of
individuals in the social space is given by $\delta =N/h_{max}$. The
distance between individuals is defined as $d(h_i,h_j)\equiv|h_i-h_j|$.
Therefore, the controlling parameter in the model is the homophyly
parameter $\alpha$.
\begin{figure}
\begin{center}
\begin{tabular}{ccc}
   & $\alpha=1.1$ &  \\
  \epsfig{file=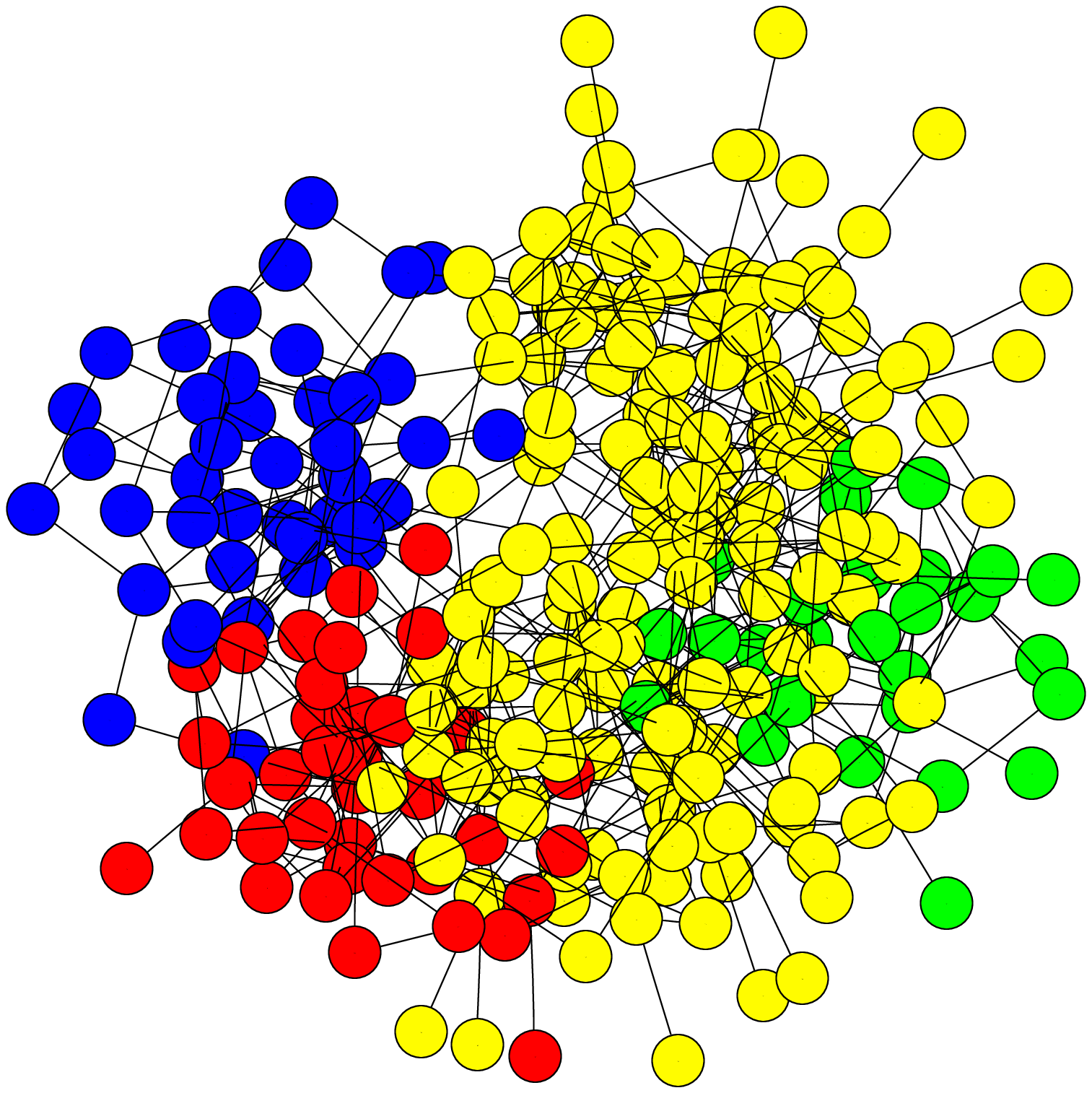, width=3cm} & &\epsfig{file=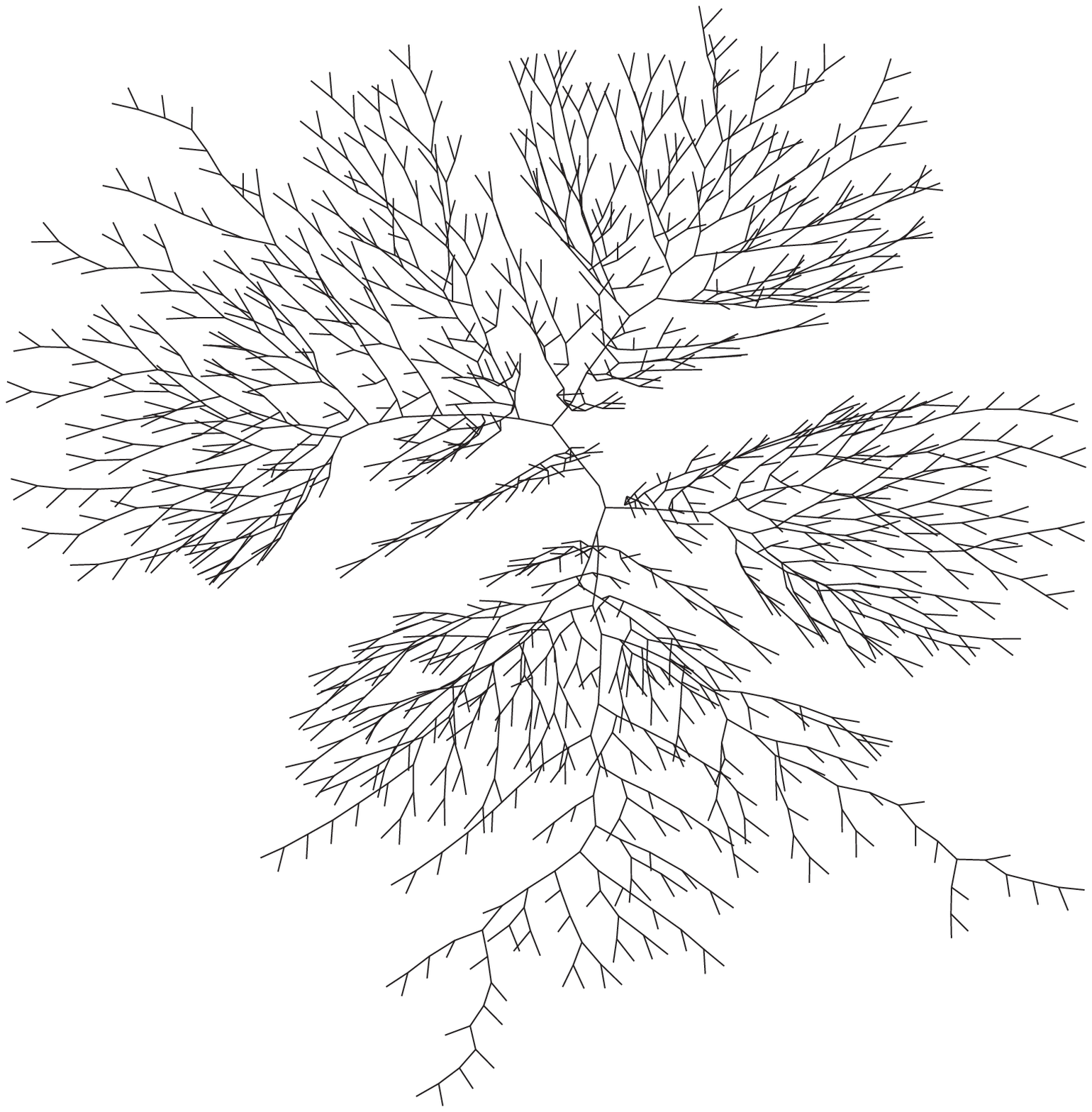, width=3cm} \\
   & $\alpha=2$ &  \\
  \epsfig{file=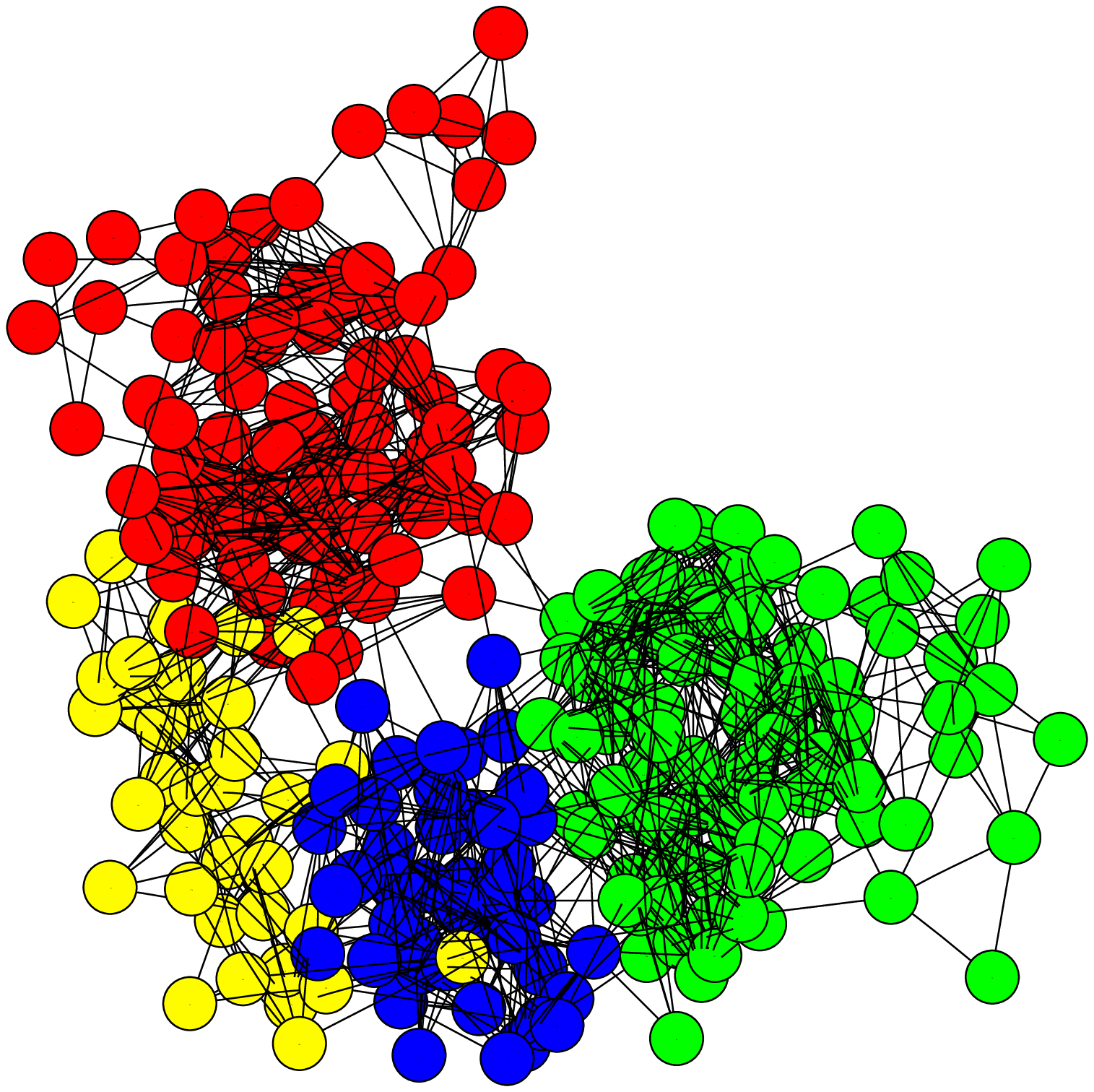, width=3cm}&   & \epsfig{file=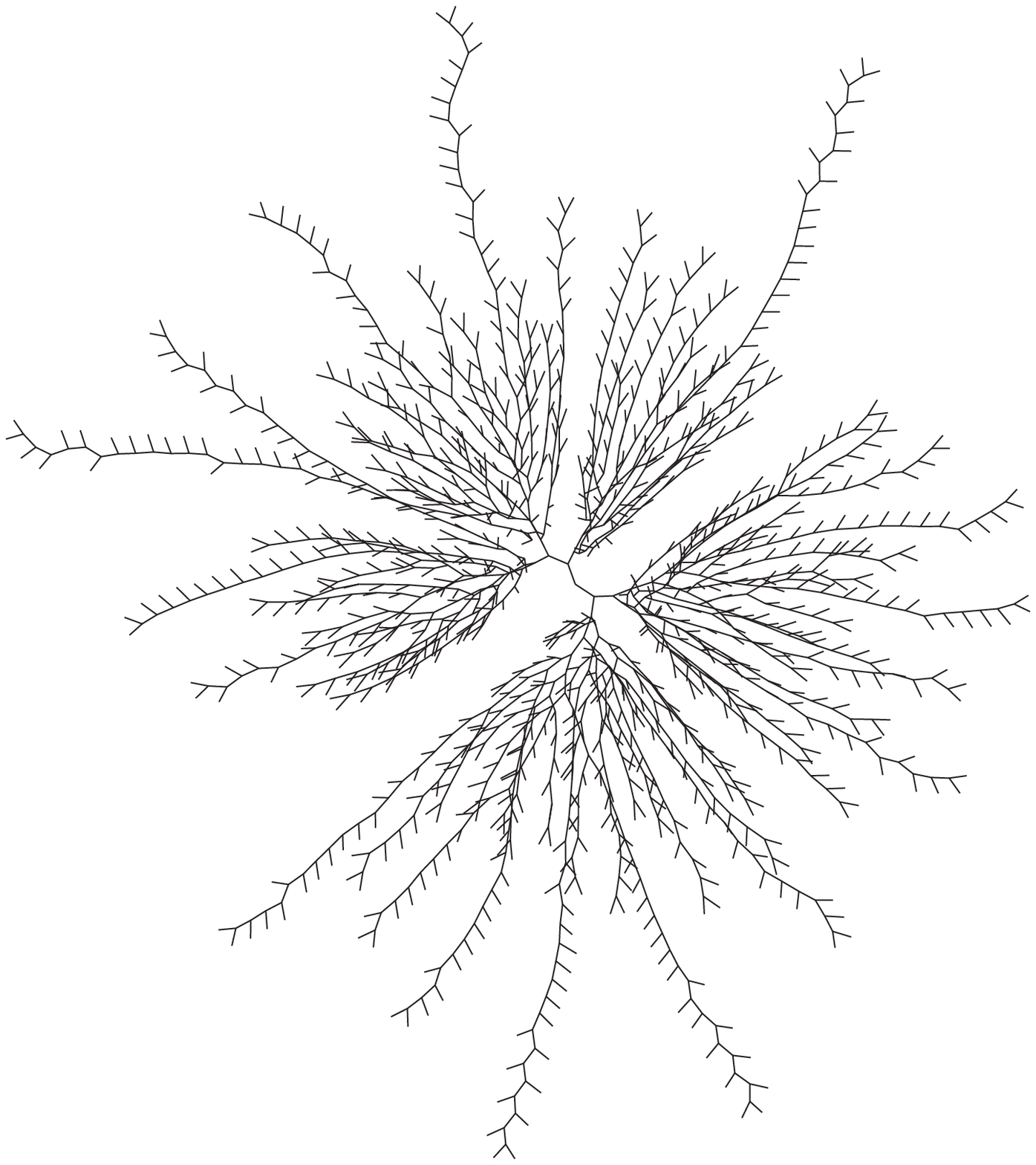, width=3cm} \\
   & $\alpha=3$ &  \\
  \epsfig{file=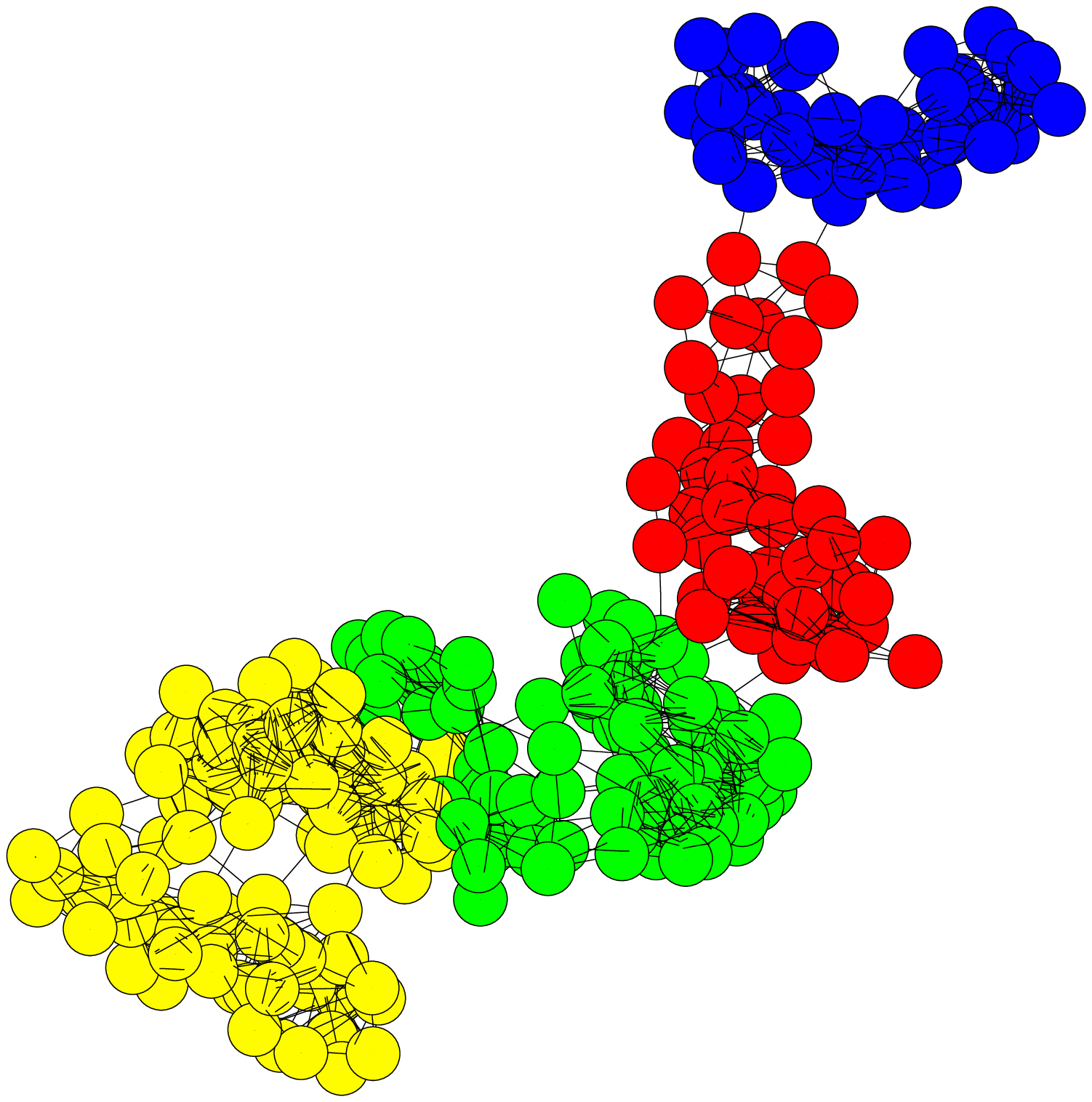, width=3cm} & & \epsfig{file=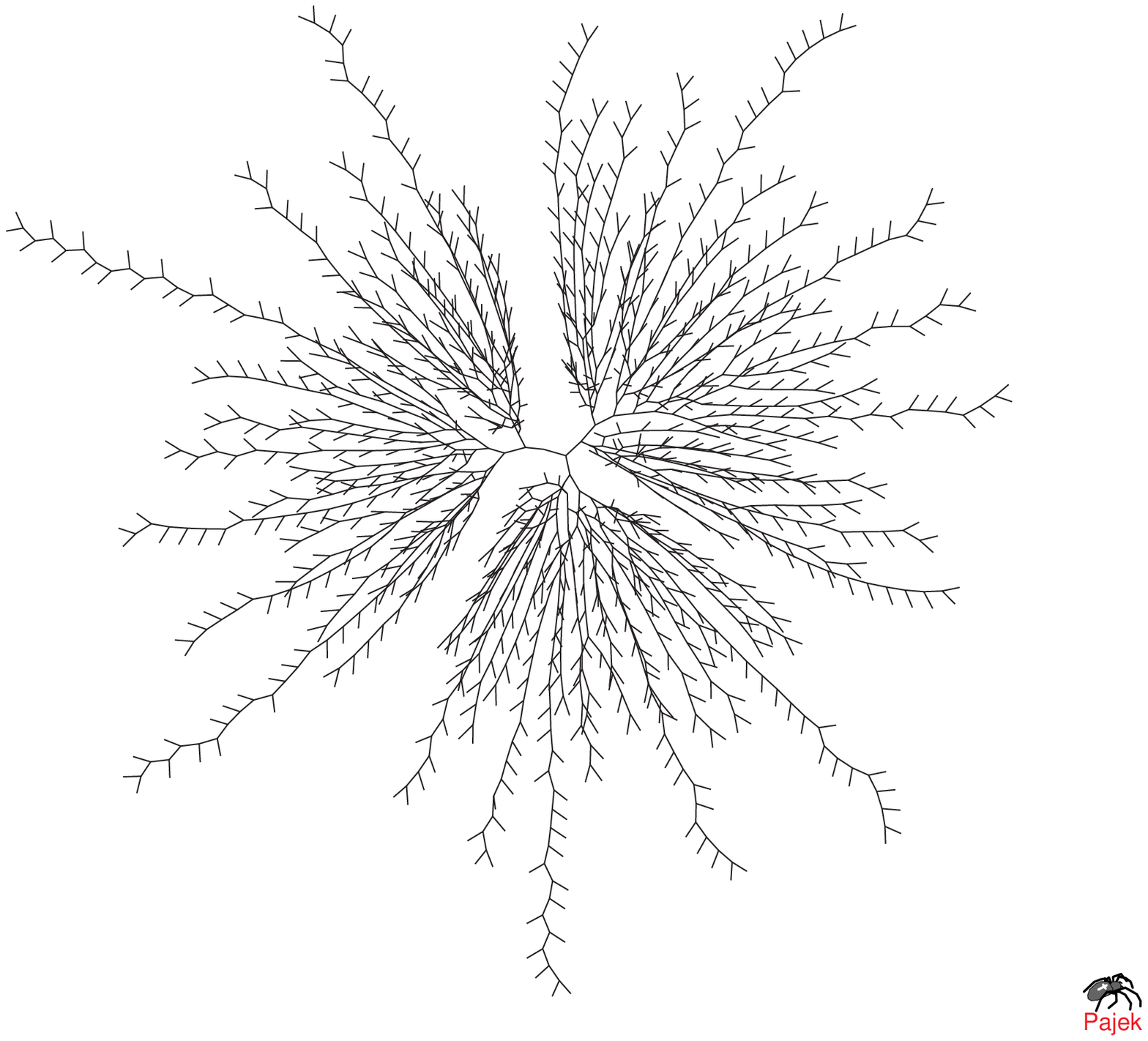, width=3cm}
\end{tabular}
\end{center}
\caption{Left: Examples of typical networks generated for an average
  degree $\langle k \rangle =10$, $N=250$, $\delta =2$, and different values of the
  parameter $\alpha$. Right: Binary trees representing the community
  structure of the corresponding networks (see text).}
\label{nets&trees}
\end{figure}
The left panel of Fig.~\ref{nets&trees} shows some typical examples of
networks generated with our model, for different values of the
parameter $\alpha$.

The model, as defined above, is homogeneous in the limit $h_{max} \gg
1$, which means that all the vertex properties will eventually become
independent of the social coordinate $h$. Therefore, the average
degree can be calculated as $\langle k \rangle =\lim_{h_{max}\to \infty}
\bar{k}(h=h_{max}/2)$ which leads to
\begin{equation}
\langle k \rangle=\frac{2 \delta  b\pi}{\alpha \sin{\pi/\alpha}}.
\label{avdegree}
\end{equation}
Thus, for fixed $\delta$, we can construct networks with the same
average degree and different homophyly, $\alpha$, by changing $b$
according to the previous expression. For $\alpha=1$ the average
degree diverges because, in this case, there is a finite
probability of connection to infinitely distant vertices.
\begin{figure}
\epsfig{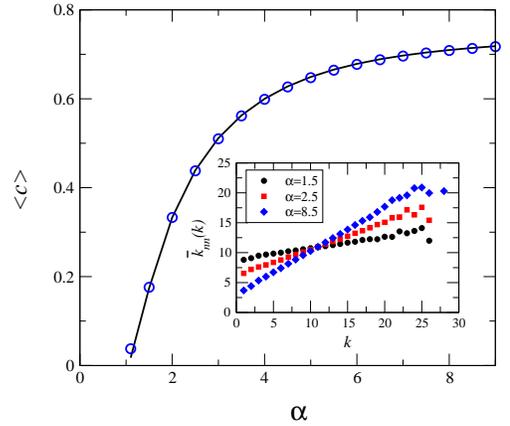}
\caption{Clustering coefficient for $d_\mathcal{H}=1$ as a
  function of $\alpha$ and fixed average degree, $\langle k \rangle =10$. The solid
  line corresponds to the theoretical value
  Eq.~(\ref{clustering_teoric}) and symbols are simulation results.
  Inset: Average nearest neighbors degree for $d_\mathcal{H}=1$ as a
  function of $k$, for different values of $\alpha$. In all cases,
  the size of the network is $N=10^5$.}
 \label{clus-knn}
\end{figure}
The clustering coefficient can be computed by means of
Eq.~(\ref{cluster_h}), yielding
\begin{equation}
  \langle c \rangle=\frac{\alpha^2}{4 \pi^2} f(\alpha) \sin^2{\frac{\pi}{\alpha}}
\label{clustering_teoric}
\end{equation}
where
\begin{equation}
f(\alpha)= \int_{-\infty}^{\infty} \int_{-\infty}^{\infty}
\frac{dx dy}{(1+|x|^{\alpha})(1+|x-y|^{\alpha})(1+|y|^{\alpha})}
\end{equation}
Fig.~\ref{clus-knn} shows the perfect agreement between simulations of
the model compared to the theoretic value
Eq.~(\ref{clustering_teoric}), computed by numerical integration.  We
observe that the clustering coefficient vanishes when $\alpha =1$, that
is, for weakly homophyllic societies, and converges to a constant
value $\langle c \rangle =3/4$ when $\alpha \to \infty$~\footnote{$f(\alpha \to \infty)$ can be
  exactly computed by noticing that the functions within the integrals
  approach, in this limit, to step functions.}, which corresponds to a
strongly homophyllic society.

Regarding the degree correlations, at first sight one could conclude
that, since the network is homogeneous in the social space
$\mathcal{H}$, the resulting network is free of any correlations.
However, numerical simulations of the average degree of the nearest
neighbors as a function of the degree, $\bar{k}_{nn}(k)$, show a
linear dependence on $k$ and, consequently, assortative mixing by
degree (see Fig.~\ref{clus-knn}).  This counterintuitive result is a
consequence of the fluctuations of the density of individuals in the
social space.  Indeed, if individuals are placed in the space
$\mathcal{H}$ with some type of randomness, they will end up forming
clusters (communities) of close individuals, strongly connected among
them.  Therefore, an individual with large degree will most probably
belong to a large cluster, and consequently its neighbors will have
also a high degree.

\begin{figure}
  \epsfig{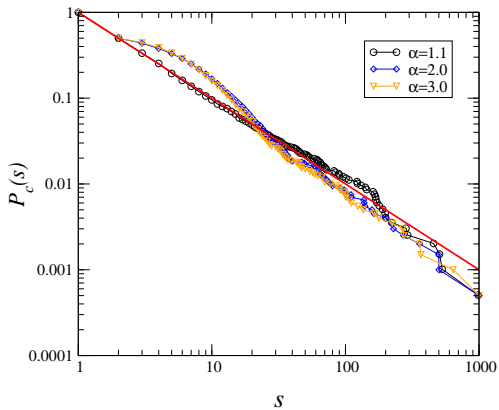}
  \caption{Cumulative size
    distribution obtained using the GN algorithm for values of
    $\alpha=1.1$, $2$ and $3$. As $\alpha \to 1$ the network becomes a
    perfectly hierarchical network characterized by a power law
    community size distribution, $P(s)\sim s^{-2}$. In all the cases the
    size of the network is $N=1000$.} \label{sizedist}
\end{figure}

Finally, we focus on the community structure displayed by our model.
To this purpose, we use the algorithm proposed by Girvan and Newman
(GN) \cite{girvan02} to identify communities in complex networks. The
performance of this algorithm relies on the fact that edges connecting
different communities have high betweenness (a centrality measure of
vertex and edges of the network \cite{freeman77}, that is defined as
the total number of shortest paths among pairs of vertices of the
network that pass through a given vertex or edge \cite{brandesbet}).
The algorithm recursively identifies and cuts the edge with the
highest betweenness, splitting the network until the single vertex
level. The information of the entire process can be encoded into the
binary tree generated by the splitting procedure. The advantage of
using the binary tree representation is twofold, since it gives
information about the different communities---which are the branches
of the tree---and, at the same time, unravels the hierarchy of such
communities.  The right panel of Fig.~\ref{nets&trees} shows the
binary trees corresponding to the networks shown in the left panel.
As $\alpha$ grows, the network eventually becomes a chain of clusters
connected by a few edges. In contrast, as $\alpha$ approaches $1$ the
network is more and more interconnected and develops a hierarchical
structure. This hierarchical structure can be quantified by means of
the cumulative distribution of community sizes, $P_c(s)$, in which the
community size $s$ is defined as the number of individuals belonging
to each offspring during the splitting procedure.  Fig.~\ref{sizedist}
shows $P_c(s)$ for $\alpha=1.1$, $2$ and $3$. When $\alpha \sim 1$, the
cumulative size distribution approaches to $P_c(s) \sim s^{-1}$,
reflecting the hierarchical structure of the network.  For higher
values of $\alpha$ the hierarchy is still preserved for large community
sizes whereas for small sizes there is a clear deviation as a
consequence of clusters of highly connected individuals which form
indivisible communities, breaking thus the hierarchical structure at
low levels. These clusters are identified in the binary tree as the
long branches with many leaves at the end of the tree.

To sum up, in this paper we have presented a model of social network
with non-zero clustering coefficient in the thermodynamic limit,
assortative degree mixing, and a hierarchical (self-similar) community
structure. The origin of these properties can be traced back to the
very presence of communities, due to the fluctuations in the position
of individuals in social space. Our approach opens thus new views for
a further understanding of the structure of complex social networks.

\begin{acknowledgments}
This work has been supported by DGES of the Spanish Government,
Grant No. BFM2000-0626 and EC-FET Open Project No. IST-2001-33555.
R.P.-S. acknowledges financial support from the MCyT (Spain), and
from the DURSI, Generalitat de Catalunya (Spain). We thank R.
Guimer{\`a}, L.A.N. Amaral, and A. Vespignani for helpful discussions.

\end{acknowledgments}

\end{document}